\documentclass[12pt]{article}
\usepackage{graphicx}
\makeatletter

\def\paragraph{\@startsection{paragraph}{4}{\z@}{+2.00ex plus
 +1ex minus +.2ex}{1.5ex plus .2ex}{\it\normalsize}}

\def\section{\@startsection {section}{1}{\z@}{+3.0ex plus +1ex minus
  +.2ex}{2.3ex plus .2ex}{\normalsize\bf\boldmath}}
\def\subsection{\@startsection{subsection}{2}{\z@}{+2.5ex plus +1ex
minus +.2ex}{1.5ex plus .2ex}{\normalsize\bf\boldmath}}
\def\subsubsection{\@startsection{subsubsection}{3}{\z@}{+3.25ex plus
 +1ex minus +.2ex}{1.5ex plus .2ex}{\normalsize\bf\boldmath}}

\expandafter\ifx\csname mathrm\endcsname\relax\def\mathrm#1{{\rm #1}}\fi

\newcounter{saveeqn}

\@addtoreset{equation}{section}

\begin{document}
\begin{titlepage}
\noindent February 28, 2003      \hfill    LBNL-52452\\

\begin{center}

\vskip .5in

{\large \bf The Direct Limit on the Higgs Mass and the SM Fit}
\footnote
{This work is supported in part by the Director, Office of Science, Office
of High Energy and Nuclear Physics, Division of High Energy Physics, of the
U.S. Department of Energy under Contract DE-AC03-76SF00098}

\vskip 10pt
{\it Presented at the Workshop on Electroweak Precision Data and the 
Higgs Mass \\
DESY-Zeuthen, February 28 - March 1, 2003\\
To be published in the proceedings}
\vskip .5in

Michael S. Chanowitz\footnote{Email: chanowitz@lbl.gov}

\vskip .2in

{\em Theoretical Physics Group\\
     Ernest Orlando Lawrence Berkeley National Laboratory\\
     University of California\\
     Berkeley, California 94720}
\end{center}

\vskip .25in

\begin{abstract}

Because of two $3\sigma$ anomalies, the Standard Model (SM) fit of the
precision electroweak data has a poor confidence level, $CL= 0.02$.
Since both anomalies involve challenging systematic issues, it might
appear that the SM could still be valid if the anomalies resulted from
underestimated systematic error. Indeed the $CL$ of the global fit
could then increase to 0.71, but that fit predicts a small Higgs boson
mass, $m_H=45$ GeV, that is inconsistent at 95\% CL with the lower
limit, $m_H>114$ GeV, established by direct searches.  The data then
favor new physics if the anomalous measurements are both excluded or
both retained, and the Higgs boson mass cannot be predicted until the
new physics is understood. The validity of the SM could however be
maintained by a propitious combination of statistical fluctuation and
systematic error. The current data do not allow a definitive
conclusion.

\end{abstract}

\end{titlepage}

\renewcommand{\thepage}{\roman{page}}
\setcounter{page}{2}
\mbox{ }

\vskip 1in

\begin{center}
{\bf Disclaimer}
\end{center}

\vskip .2in

\begin{scriptsize}
\begin{quotation}
This document was prepared as an account of work sponsored by the United
States Government. While this document is believed to contain correct
 information, neither the United States Government nor any agency
thereof, nor The Regents of the University of California, nor any of their
employees, makes any warranty, express or implied, or assumes any legal
liability or responsibility for the accuracy, completeness, or usefulness
of any information, apparatus, product, or process disclosed, or represents
that its use would not infringe privately owned rights.  Reference herein
to any specific commercial products process, or service by its trade name,
trademark, manufacturer, or otherwise, does not necessarily constitute or
imply its endorsement, recommendation, or favoring by the United States
Government or any agency thereof, or The Regents of the University of
California.  The views and opinions of authors expressed herein do not
necessarily state or reflect those of the United States Government or any
agency thereof, or The Regents of the University of California.
\end{quotation}
\end{scriptsize}

\vskip 2in

\begin{center}
\begin{small}
{\it Lawrence Berkeley National Laboratory is an equal opportunity employer.}
\end{small}
\end{center}

\newpage

\renewcommand{\thepage}{\arabic{page}}
\setcounter{page}{1}

\vspace*{1cm}
\begin{center}

{\Large \bf The Direct Limit on the Higgs Mass and the SM Fit}

\vspace*{1cm}

{\sc Michael S. Chanowitz}

\vspace*{.5cm}

{\normalsize \it
Lawrence Berkeley National Laboratory \\
University of California\\
Berkeley, California 94720}
\par
\end{center}
\vskip 1cm
\begin{center}
\bf Abstract
\end{center} 
{\it
Because of two $3\sigma$ anomalies, the Standard Model (SM) fit of the
precision electroweak data has a poor confidence level, $CL= 0.02$.
Since both anomalies involve challenging systematic issues, it might
appear that the SM could still be valid if the anomalies resulted from
underestimated systematic error. Indeed the $CL$ of the global fit
could then increase to 0.71, but that fit predicts a small Higgs boson
mass, $m_H=45$ GeV, that is inconsistent at 95\% CL with the lower
limit, $m_H>114$ GeV, established by direct searches.  The data then
favor new physics if the anomalous measurements are both excluded or
both retained, and the Higgs boson mass cannot be predicted until the
new physics is understood. The validity of the SM could however be
maintained by a propitious combination of statistical fluctuation and
systematic error. The current data do not allow a definitive
conclusion.
}
\par
\vskip 1cm

\section{Introduction}

A decade of beautiful experiments at CERN, Fermilab, and SLAC have
provided increasingly precise tests of the Standard Model (SM). The
data confirms the SM at the level of quantum effects and probes the
Higgs boson mass. The global fit\footnote{We fit only the observables
considered by the EWWG prior to 2002, including neither APV
nor $\Gamma_W$. Theoretical systematics are not clearly controlled for
the former while the latter as a 3\% measurement is insensitive to 
SM-level radiative corrections. 
Including them the fit would yield
$CL=0.04$.}  has a poor confidence level $CL=0.02$, due to two
$3\sigma$ anomalies.  One of these, $x_W^{\rm OS}[ {\nu} N]$ from
NUTEV\cite{mcfarland}, is recent.  The other, the discrepancy between
the effective leptonic mixing angle, $x_W^l = {\rm sin}^2\theta^l_W$,
determined from three leptonic asymmetry measurements($x_W^l[A_L]$)
versus its determination from three hadronic ($x_W^l[A_H]$)
measurements, is dominated by the $\sim 3\sigma$ discrepancy between
the two most precise, $A_{LR}$\cite{swartz} and
$A_{FB}^b$,\cite{wells} which has been a persistent feature of the
data since the earliest days of LEP and SLC.

In this talk I focus on how the direct lower limit from LEP II, $m_H >
114.4$ GeV,\footnote { N.B., the 95\% lower limit from
the direct searches does {\em not} imply a 5\% chance that $m_H <
114.4$ but rather means that if the mass were actually 114.4 GeV it
could have escaped detection with 5\% likelihood. The likelihood for
$m_H < 114.4$ GeV is $\ll 5\%$. See for instance section 5 of
\cite{mchvr}.} constrains the interpretation of these anomalies, and
especially the asymmetry anomaly.\cite{msc} The central observation is
that the only measurements which support $m_H$ in the allowed region
are precisely the ones with big pulls that drive the fit to a poor
CL. Without the discrepant measurements the prediction for $m_H$ is
too low.  In particular, the agreement of the SM with the data would
not be improved if both anomalies were attributed to systematic error,
since the resulting fit (with $x_W^{\rm OS}[{\nu} N]$ and $x_W^l[A_H]$
removed) predicts $m_H=45$ GeV, with only a 5\% likelihood that $m_H >
114$ GeV.

The $W$ mass measurement plays a central role, discussed in more
detail in \cite{msc}. In the SM fit, it favors the lower range of the
leptonic asymmetries, $x_W^l[A_L]= 0.23113(21)$, over the larger
hadronic result, $x_W^l[A_H]= 0.23217(29)$.  Using the two loop result
of \cite{fhw} the new experimental result,
$m_W=80.426(34)$GeV,\cite{hawkings,grunewald} implies $x_W^l=0.23095$
and a very light value for $m_H$.  Essentially it is $m_W$ which
decides whether $A_{LR}$ or $A_{FB}^b$ will have the largest pull. 

The interpretation of the data is not clear. All three generic
possibilities are in play: new physics, statistical fluctuation, and
underestimated systematic error. It is certainly possible that either
anomaly is genuine evidence of new physics,\cite{newphys1} in which
case the SM fit would be invalidated and we could not use the
precision data to constrain the Higgs boson mass until the new physics
were understood.  Statistical fluctation is also a possible
explanation, which is fairly represented by the global CL's. The new
23 MeV downward shift in $m_W$ increases the global CL by a factor
two, from 0.01 to 0.02, which still cannot be said to be ``favored''.

Concerning the possibility of underestimated systematic error, the
three leptonic asymmetry measurements, $x[A_L]$, are theoretically
clean and use three quite different experimental techniques, so 
that a large common systematic error is very unlikely. In
contrast, both $x_W^{\rm OS}[{\nu} N]$ and $x[A_H]$ depend on subtle
systematic issues, involving experimental technique and, especially,
nontrivial applications of QCD. The three hadronic asymmetry
measurements have important shared systematics, both theoretical and
experimental.  If the systematic uncertainties of the $x_W^{\rm
OS}[{\nu} N]$ and $x[A_H]$ anomalies were much larger than current
estimates, the $CL$ of the global fit could increase to as much as
0.71. The SM might then appear to provide a good description of the
data, however we would then encounter the conflict with the LEP II
lower limit on $m_H$.  This conflict would also signify new
physics,\cite{newphys2} to raise the prediction for $m_H$ into the
allowed region above 114 GeV. Again $m_H$ could not be predicted until
the new physics is known. With oblique corrections it is possible to
``dial in'' essentially any value of $m_H$.\cite{msc}

It should be clear that the focus here on the possibility of underestimated
systematic error is not based on the belief that it is the most likely
explanation. In both cases the experimental groups have put great
effort into understanding and estimating the systematic uncertainties,
and the quoted systematic errors are too small to explain the
anomalies.\cite{mcfarland,wells} In fact, the situation is truly
puzzling, and there is no decisive reason to prefer systematic error
over new physics as the explanation of either anomaly. Rather we
consider the systematic error hypothesis simply in order to understand
what it implies and find that it also points to new physics.

Though it is an {\it a posteriori} observation, the grouping of the
six asymmetry measurements into hadronic and leptonic clusters is a
striking feature of the data.  The leptonic asymmetries are the three
lowest, combining to $x[A_L]=0.23113(21)$ with $\chi^2/dof = 1.7/2$,
$CL= 0.43$. The hadronic asymmetries are the three highest, tightly
clustered around $x[A_H]=0.23217(29)$, with $\chi^2/dof = 0.05/2$,
$CL= 0.97$. Combining all six measurements we have $x_W^l=0.23148(17)$ with
$\chi^2/dof = 10.2/5$ and $CL=0.07$.  It is unclear whether the
grouping into leptonic and hadronic clusters is by chance or whether
it is telling us something, either about new physics or about
systematic effects. Since they are linked by common systematics, we
consider the the three hadronic asymmetry measurements together when
considering the systematic uncertainty hypothesis.

\section{\bf SM Fits}

The SM radiative corrections are computed with ZFITTER
v6.30,\cite{zfitter} but with the two loop $m_W$.\cite{fhw}
Experimental correlations are from \cite{ewwg_02} and $\Delta
\alpha_5(m_Z)$ is from \cite{bp} as in \cite{ewwg_02}. Predictions for
$m_H$ are obtained using $\Delta \chi^2$\cite{ewwg_02} and also
with a ``Bayesian'' likelihood method\cite{msc}. Both methods give
very similar results, and only the former are reported here. When
fitting the same observables the SM predictions and $\chi^2$ results
agree well with \cite{grunewald,ewwg_02}, with small differences from
our use of \cite{fhw}.

Table 1 summarizes $\chi^2$ fits of four data sets, in which none,
one, or both sets of anomalous measurements are excluded. We vary
$m_t$ and $\Delta \alpha_5(m_Z)$, which are constrained, and
$\alpha_S(m_Z)$ and $m_H$, which are unconstrained.  Fit A is our
``all data''set, including the ten $m_H$-sensitive observables (the
six $x_W^l$ determinations, $x_W^{\rm OS}[{\nu} N]$, and three
``non-asymmetry'' observables, $m_W$, $\Gamma_Z$, and $R_l$) and 
five $m_H$-insensitive observables ($\sigma_h$, $R_b$, $R_c$, $A_b$,
and $A_c$). The CL increases from 0.02 in fit A to 0.17 if $x_W^{\rm
OS}[{\nu} N]$ is omitted (B), or to 0.08 if the hadronic asymmetries
are omitted (C), to 0.71 if both are omitted (D).

Fits restricted to the $m_H$-sensitive sector, which determines the SM
prediction for $m_H$, are also shown in table 1. Fit A$^\prime$, with
all ten $m_H$-sensitive observables, has $CL=0.005$, while B$^\prime$
and C$^\prime$ are at 0.07 and 0.02 respectively, each substantially
smaller than the corresponding global fits A,B,C.  A fit of just the
three most precise $m_H$-sensitive observables, $A_{LR}$, $A_{FB}^b$,
and $m_W$, which together dominate the $m_H$ prediction,\footnote{The
fit to these three observables has $m_H=94$ with $36<m_H<212$ (90\%
CL), compared with $m_H=90$ and $39<m_H<205$ from the all-data global
fit A.}  yields $\chi^2/dof=10.2/2$ and $CL=0.006$. The poor
consistency of the $m_H$-sensitive sector is cause for concern in
assessing the reliability of the SM prediction of $m_H$.

\vskip 12pt
\noindent {\bf Table 1.} Results for global fits A - D and for the
corresponding fits restricted to $m_H$-sensitive observables, 
A$^\prime$ - D$^\prime$.

\begin{center}
\begin{tabular}{c||c|c}
 & All &$- x_W^{\rm OS}[{\nu} N]$ \\
\hline
\hline
  All & \bf A & \bf B \\
    & $\chi^2/=25.7/13$, $CL=0.019$ & 16.5/12, 0.17 \\
\hline
  $-x_W^l[A_H]$ & {\bf C} & \bf D \\
                & 16.7/10, 0.081  &  6.3/9, 0.71 \\
\hline
\hline
$m_H$-sensitive only: & &  \\
All & \bf A$^\prime$ & \bf B$^\prime$ \\
    & 22.2/8, 0.0046 &  13.2/7, 0.067 \\
\hline 
  $-x_W^l[A_H]$ & {\bf C$^\prime$} & \bf D$^\prime$ \\
    & 13.2/5, 0.022 & 2.94/4, 0.57 \\
\hline
\hline
\end{tabular}
\end{center}

\begin{figure}[t]
\begin{center}
\rotatebox{90}{\includegraphics[width= .5\textwidth]{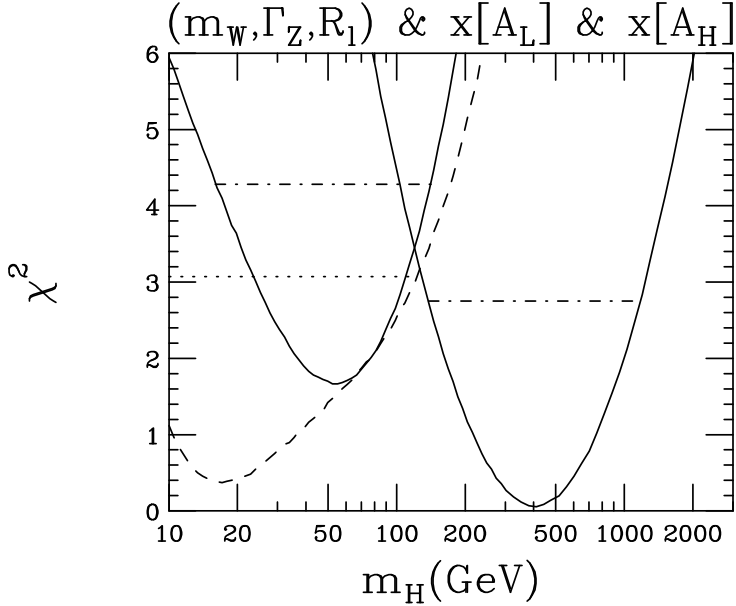}}
\caption{$\chi^2$ distributions for $x_l^W[A_L]$ and $x_l^W[A_H]$ 
(solid lines) and for $(m_W,\Gamma_Z,R_l)$ (dashed line). The symmetric 
90\% CL intervals are indicated by the dot-dashed lines for 
$x_l^W[A_L]$ and $x_l^W[A_H]$ and the dotted line for $(m_W,\Gamma_Z,R_l)$.
}
\end{center}
\end{figure}

The global CL's fairly reflect the likelihoods of the fits. Consider
for example fit B, for which $A_{FB}^b$ is the only significant
outlyer, with a pull of 2.59. While $2.59\sigma$ corresponds to $CL=
0.0096$, in the context of fit B we should ask for the probability
that at least one of 12 independent measurements would deviate by
$\geq 2.59\sigma$. This is $1 - (1-0.0096)^{12}=0.11$, which
appropriately reflects $CL=0.17$ from the $\chi^2$ global fit.

It is also instructive to consider the predictions of the
$m_H$-sensitive observables, table 2. For $A_{LR}$ the central value
is $m_H=39$ GeV, the 95\% upper limit is 122 GeV, and the 

\vskip 12pt
\noindent {\bf Table 2.} Predictions for $m_H$ from the three highest
precision $m_H$-sensitive observables, from the combined leptonic
asymmetries $x[A_L]$, the combined hadronic asymmetries $x[A_H]$, and the
three remaining (non-asymmetry) $m_H$-sensitive observables. 
The value of $m_H$ at the $\chi^2$ minimum is shown along with the
symmetric 90\% confidence interval and the likelihood for $m_H>114$
GeV.  Values indicated as $10-$ or $3000+$ fall below or above the
interval $10 <m_H<3000$ GeV within which the fits are performed.
\begin{center}
\begin{tabular}{c|ccc}
 & $m_H$ (GeV) & 90\% $CL$ & $CL(m_H>114)$ \\ 
\hline
\hline
$A_{LR}$ & 39 & $10- < m_H< 122$ & 0.062 \\
$A_{FB}^b$ & 410 & $130 <m_H< 1200$ & 0.97 \\
$m_W$ & 35 & $10- <m_H< 161$ & 0.12\\
\hline
$x_W^l[A_L]$ & 55 & $16 <m_H<143$ & 0.10 \\
$x_W^l[A_H]$ & 410 & $140 <m_H< 1200$ & 0.97 \\
$m_W\oplus \Gamma_Z \oplus R_l$ & 17 &$ 10- <m_H< 123$ & 0.057 \\
\hline 
\hline
\hline
\end{tabular}
\end{center}

\noindent likelihood for $m_H>114$ GeV is $CL(m_H>114)= 0.06$. The $W$
mass also prefers small $m_H$. The only important contributor to large
$m_H$ is $A_{FB}^b$, with central value $m_H=410$ GeV and symmetric
90\% CL interval up to 1200 GeV. Also shown are the predictions of the
three leptonic asymmetries which are similar to $A_{LR}$, the three
hadronic asymmetries which are nearly identical to $A_{FB}^b$ (since
it has much greater precision than $A_{FB}^c$ and $Q_{FB}$), and the
non-asymmetry measurements $(m_W,\Gamma_Z,R_l)$ which resemble $m_W$
though with a stronger preference for light $m_H$. The $\chi^2$
distributions are shown in figure 1.

Table 3 summarizes the $m_H$ predictions of the four global fits for
which $CL(\chi^2)$ is shown in table 1. Fit A is similar to the EWWG
all-data fit\cite{grunewald}, with the 95\% CL upper limit at
$m_H<205$ GeV. The omission of $x_W^{\rm OS}[{\nu} N]$ from fit B
increases $CL(\chi^2)$ appreciably but has little effect on $m_H$.
Fits C and D, with the hadronic asymmetry measurements excluded, both
have $m_H=45$ GeV as central value, with 7\% and 5\% CL's respectively
for $m_H>114$ GeV. The $\Delta \chi^2$ distributions for fits A and D
are shown in figure 2.

Since the internal consistency of the global fit, $CL(\chi^2)$, and the
consistency of the fit with the search limit, $CL(m_H>114)$, are
independent, it is interesting to consider the combined
probability, given by the product
$$
P_C= CL(\chi^2) \times CL(m_H > 114).    \eqno{(1)}
$$ 
The {\em relative} values of $P_C$ for the different fits are
especially interesting.  From table 3 we see that $P_C$ is
approximately independent of whether the hadronic asymmetry
measurements $x_W^l[A_H]$ are included, although the individual
factors on the right side of eq. (1) are very     
sensitive to 
$x_W^l[A_H]$. For instance, for global fit A ( `all'
data), $CL(\chi^2) =0.019$ and $CL(m_H >114)= 0.35$ so that $P_C[A]=
0.019 \times 0.35= 0.0066$.For fit C, with the three $x_W^l[A_H]$
omitted, we have $CL(\chi^2)=0.081$, $CL(m_H >114)= 0.068$, and
$P_C[C]= 0.081 \times 0.068= 0.0055 \simeq P_C[A]$. Similarly,
$P_C[B]= 0.17 \times 0.29 = 0.049$ and $P_C[D]=0.71 \times 0.049 =
0.035 \simeq P_C[B]$. If the  hadronic asymmetry
measurements are omitted, the 
increase in the global fit confidence
level is compensated by a roughly equal decrease in the consistency 
with the direct search limit.

\vskip 12pt
\noindent {\bf Table 3.} Higgs boson mass predictions for global fits
A - D. Each entry shows the value of $m_H$ at the $\chi^2$ minimum,
the symmetric 90\% confidence interval, the CL for consistency with
the search limit, and the combined likelihood $P_C$, eq. (1), 
with the factors $CL(\chi^2)$ and $CL(m_H > 114)$ explicitly
displayed.

\begin{center}
\begin{tabular}{c||c|c}
 & All &$- x_W^{\rm OS}[\stackrel {(-)}{\nu} N]$ \\
\hline
\hline
  All & \bf A & \bf B \\
   & $m_H= 90$ & $m_H= 90$  \\  
   & $39<m_H<205$ & $38<m_H<190$  \\              
   & $CL(m_H>114)=0.35$ & $CL(m_H>114)=0.29$  \\
   & $P_C= 0.019 \times 0.35=0.0066$ & $P_C=0.17 \times 0.29=0.049$ \\
\hline
  $-x_W^l[A_H]$ & {\bf C} & \bf D \\
   & $m_H= 45$ & $m_H= 45$  \\              
   & $18<m_H<123$ & $18<m_H<114$  \\   
   & $CL(m_H>114)=0.068$ &  $CL(m_H>114)=0.049$  \\
   & $P_C=0.081\times 0.068=0.0055$ & $P_C= 0.71\times 0.049= 0.035$ \\
\hline
\hline
\end{tabular}
\end{center}

It is easy to show that the product of $n$ independent CL's has an
expectation value of $1/2^n$ so that the expectation value of $P_C$ is
0.25. Consider a continuous probability density function $f(x)$
defined for positive $x$, with normalization
$$
\int^{\infty}_0 dx f(x) = 1.    \eqno{(2)}
$$
Then the CL for $x < y$ is 
$$
F(y)= \int^y_0 dx f(x)         \eqno{(3)}
$$
and the expected value of F
is 
$$
<F> = \int^{\infty}_0 dy F(y) f(y) = {1 \over 2} \eqno{(4)} 
$$ 
where equations (2) and (3) suffice to evaluate the integral.  For a
product of $n$ independent CL's the expectation value of the combined
product CL is then given by $n$ factors of the form of equation (4),
with the result $1/2^n$. In the case at hand $P_C$ is the product of
two $\chi^2$ distributions: $CL(\chi^2,N)$ and $CL(m_H > 114)=
CL(\Delta \chi^2, 1)/2$, where $\Delta \chi^2$ is the difference
between the minimum at $m_H=114$ and the global minimum.  The factor
1/2 in $CL(m_H > 114)$ is compensated by the fact that the domain of
$\Delta \chi^2$ extends from $-\infty$ to $+\infty$; modifying
equations (2-4) appropriately, the expected value is again 1/2.

\begin{figure}[t]
\begin{center}
\rotatebox{90}{\includegraphics[width= .5\textwidth]{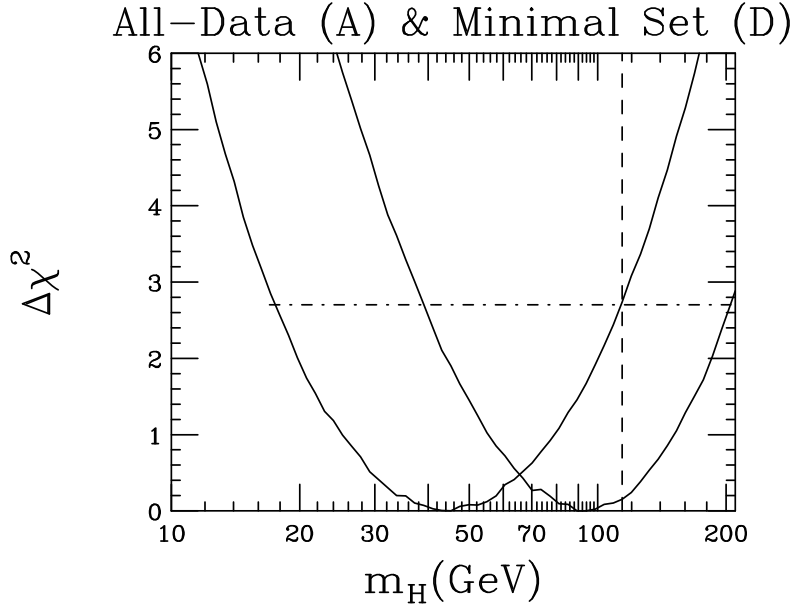}}
\caption{$\Delta \chi^2$ distributions for fits A and D. The dot-dashed 
line denotes the 95\% CL upper and lower limits and the dashed line 
indicates the experimental lower limit on $m_H$.}
\end{center}
\end{figure}

In addition to the pairings $P_C[A]\simeq P_C[C]$ and $P_C[B]\simeq
P_C[D]$, the other prominent feature of table 3 is that $CL(m_H >114)$
depends sensitively on whether $x[A_H]$ is retained but not on
$x_W^{\rm OS}[{\nu} N]$, i.e., $CL(m_H>114)_A \simeq CL(m_H>114)_B$
and $CL(m_H>114)_C \simeq CL(m_H>114)_D$.  For fits C and D, with
$x[A_H]$ omitted, the consistency with the search limit is
poor. Relative to 0.25, the value of $P_C$ is marginal for all four
fits.

\section{\bf New Physics to Increase $m_H$}

The low values of $CL(m_H > 114)$ in fits C and D are either
statistical fluctuations (e.g., of $m_t$) or they are signals of new
physics. Examples of new physics that could do the job are the MSSM
with light sneutrinos and sleptons (Altarelli {\it et al.} in
\cite{newphys2}) or a fourth generation of quarks and leptons 
with a massive neutrino (Novikov
{\it et al.} in \cite{newphys2}). An illustrative set of parameters
for the latter is $m_N \simeq 50$ GeV, $m_E \simeq 100$ GeV, $m_U
+ m_D \simeq 500$ GeV, $|m_U - m_D| \simeq 75$ GeV, and $m_H \simeq
300$ GeV.

The prediction for $m_H$ can be increased arbitrarily in models for
which the dominant effect of the new physics is via the $W$ and $Z$
boson self energies, considered in the oblique
approximation\cite{pt}. Figure 3 shows an $S,T$ fit to the minimal
data set, with the $\chi^2$ minimum read to the left and $S,T$ read to
the right.  In contrast to the SM fit with a distinct minimum at
$m_H=45$ GeV, also shown, the oblique fit is flat, with no preference
for any range of $m_H$. The confidence level is $\simeq 0.5$ and the
variation in $\chi^2$ is very small, with $\Delta \chi^2 \leq
0.2$ for $m_H \geq 20$ GeV.

\begin{figure}[h]
\begin{center}
\rotatebox{90}{\includegraphics[width= .5\textwidth]{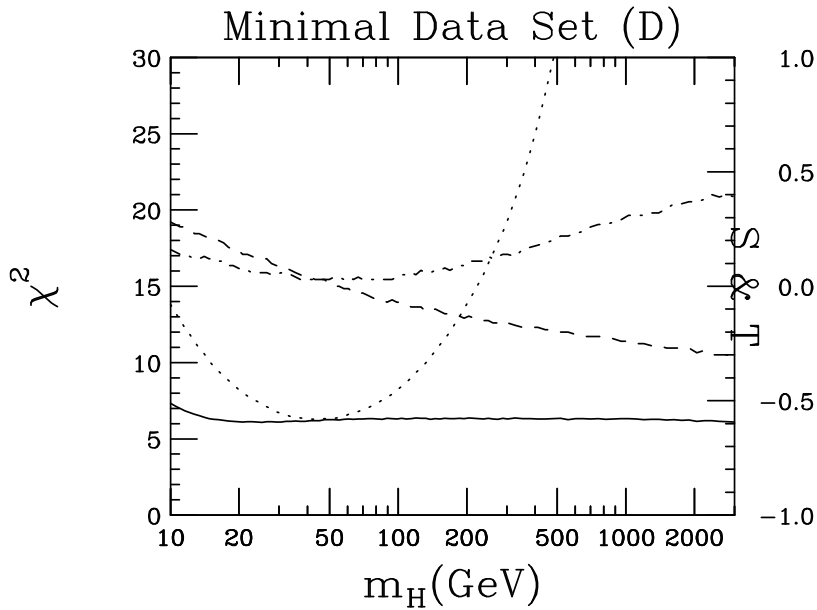}}
\caption{ Minimal Data Set (data set D) $\chi^2$ distributions for SM
(dots) and $S,T$ fit (solid). The
values of $S$ (dashed) and $T$ (dot-dashed)
are read to the right axis.}
\end{center}
\end{figure}

Values of $m_H$ above 1 TeV cannot be interpreted literally as
applying to a simple Higgs scalar. For $m_H>1$ TeV symmetry breaking
is dynamical, occurring by new strong interactions that cannot be
analyzed perturbatively.\cite{mcmkg} If the Higgs mechanism is
correct, there are new quanta that form symmetry breaking vacuum
condensates.  Values of $m_H$ above 1 TeV can be regarded only as a
rough guide to the order of magnitude of the masses of the
condensate-forming quanta.

The range of positive $T$ needed in figure 3 occurs in models with
custodial $SU(2)$ breaking, e.g., from nondegenerate quark or lepton
isospin doublets.  Negative $S$ is less natural but there is not a
no-go theorem, and models with $S<0$ have been exhibited. It is also
possible to fit the data by varying $T$ with $S=0$ fixed --- see
figure 11 of \cite{msc}.  Moderately large, postive $T$ is again
preferred. In this fit the confidence level for $m_H$ above the LEP II
lower limit is $CL(m_H>114\: {\rm GeV})= 0.21$, and the 95\% upper
limit extends to $m_H < 400$ GeV.

\section{\bf Discussion}

Taken together the precision electroweak data and the direct searches
for the Higgs boson create a complex puzzle with many possible
outcomes. An overview is given in the ``electroweak schematic
diagram,'' figure 4. The diagram illustrates how various hypotheses
about the two $3\sigma$ anomalies lead to new physics or to the
conventional SM fit. 
The principal conclusion is reflected in the fact
that the only lines leading into the `SM' box are labeled `Statistical
Fluctuation.' That is, systematic error of both $x[A_H]$ and $x_W^{OS}[\nu N]$  
cannot save the SM fit,
since it implies the conflict with the search limit, indicated by the
box labeled $CL(m_H>114)=0.05$, which in turn either implies new
physics or reflects statistical 
\begin{figure}[b!]
\begin{center}
\includegraphics[width=6.5in]{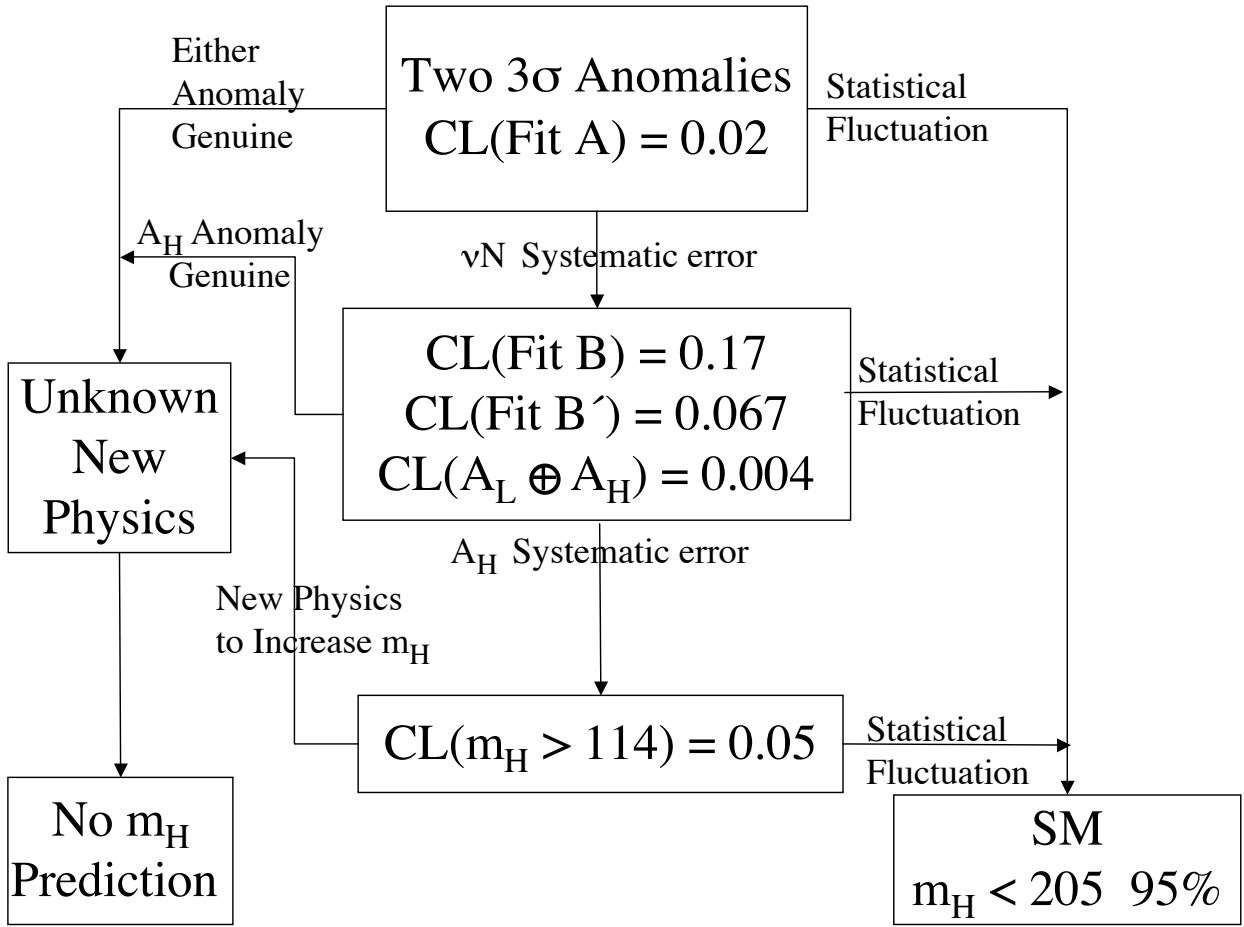}
\caption{Electroweak schematic diagram.}
\end{center}
\end{figure}
fluctuation (e.g., of
$m_t$). The problem is exhibited by the small value of $P_C$
for each of the fits in table 3.

The `New Physics' box in figure 4 is reached if either $3\sigma$
anomaly is genuine or, conversely, if neither is genuine and the
resulting 95\% $CL$ conflict with the search limit is
genuine.  
The global confidence levels of fits A and B fairly
reflect the probability that they are due to statistical
fluctuations. They do not favor the SM
and they also do not exclude it: ``It is a part of
probability that many improbable things will happen.''\cite{agathon}

The smoothest path to the SM traverses the
central box, fit B, and then exits via `Statistical Fluctuation' to
the SM. In this scenario QCD effects might explain the
NuTeV anomaly and the 17\% confidence level of fit B could be a statistical 
fluctuation. This is a valid possibility, but two other problems
indicated in the central box should also be considered in this
scenario. First, the consistency of the $m_H$-sensitive measurements
is marginal, indicated by the 6.7\% confidence level of fit
B$^\prime$. Second, the persistent conflict between the leptonic and
hadronic asymmetry measurements, currently $2.9\sigma$ with $CL=
0.0037$, is at the heart of the determination of $m_H$. Thus even if
we assume that the $CL$ of the global fit is a statistical
fluctuation, the reliability of the prediction of $m_H$ depends on
even less probable fluctuations.

The leptonic asymmetry measurements have been finalized. There are
still some ongoing analyses of the hadronic asymmetry data, but unless
major new systematic effects are uncovered, large changes are
unlikely.  To do better we will need a
second generation Z factory, such as the proposed Giga-Z project.
However, to fully exploit the potential of such a facility it will be
necessary to improve the precision of $\Delta \alpha_5(m_Z)$ by a
factor of $\sim 5$ or better, requiring a dedicated program to measure
$\sigma(e^+e^- \rightarrow {\rm HADRONS})$ below $\sim 5$ GeV to
$\simeq 1\%$.\cite{jegerlehner} The $W$ boson and top quark mass
measurements will be improved at the TeVatron, LHC, and, eventually,
at a linear $e^+e^-$ collider. 

The issues raised by the data heighten the excitement of the moment
in high energy physics.  If both $3\sigma$ anomalies reflect
systematic effects, the resulting SM fit is inconsistent with the LEP
II limit. For the SM prediction of $m_H$ to be valid the anomalies must be
a propitious combination of systematic effect ($x_W^{\nu
N}$) and statistical fluctuation ($x[A_L]$ vs. $x[A_H]$).  The end of
the decade of precision electroweak measurements leaves us with a
great puzzle, that puts into question the mass scale at which the physics of
electroweak symmetry breaking will be found.  The solution of the
puzzle could emerge at the TeVatron.  If it is not found
there it will emerge at the LHC, which at its design
luminosity will be able to search for the new quanta of the symmetry
breaking sector over the full range allowed by unitarity.\cite{mcmkg,ww}

\vskip 5pt
\noindent {\bf Acknowledgement:} I wish to thank the organizers for the 
opportunity to attend this very interesting workshop and Peter Zerwas for 
kind hospitality beyond the call of duty.


\end{document}